\documentclass[prl,aps,twocolumn,showpacs]{revtex4}
\usepackage{graphicx}

\begin{document}

\title{Interplay of density and phase fluctuations in ultracold one-dimensional Bose gases}
\author{
N.P. Proukakis}
\affiliation{Institute for Theoretical Physics, Utrecht
University, Leuvenlaan 4, 3584 CE Utrecht, The Netherlands}
%

\begin{abstract}

The relative importance of density and phase fluctuations in  ultracold
one-dimensional atomic Bose gases is investigated.
By defining appropriate characteristic temperatures for their respective onset, 
a broad experimental 
regime is found, where {\em density} fluctuations
set in at a {\em lower} temperature than
phase fluctuations. This is in stark contrast to the usual experimental regime explored up to now,
in which phase fluctuations are largely decoupled from density fluctuations, a regime 
also recovered in this work as a limiting case.
Observation of the novel regime of dominant density fluctuations
is shown to be well within current experimental capabilities
for both $^{23}Na$ and $^{87}Rb$,
requiring
relatively low temperatures, small atom numbers and moderate aspect ratios.

\end{abstract}


\maketitle

Coherence plays a crucial role in the behavior of a system, 
and is central to our understanding of laser, matter wave and condensed-matter physics.
The achievement of quantum degeneracy in ultracold one-dimensional (1D)
atomic clouds either in magnetic \cite{lowd_exp}, or optical traps \cite{lowd_exp_opt}, or 
maintained close to microfabricated surfaces known as `atom chips' \cite{AtomChip} 
has led to a unique experimentally controlable system 
for fundamental studies of coherence in low dimensional systems. Potential
applications of such systems include
 `on-chip' atom interferometers \cite{Interf,Chip_Interf}, and continuously-operating atom lasers
\cite{AL}.
1D systems differ from their 3D counterparts, in that 
low-dimensional systems are generally prone to large phase fluctuations, which tend to destroy long-range 
phase coherence \cite{popov,mullin,1d,MFT,Luxat,More}.
Nonetheless, harmonically-confined 1D gases can maintain significant coherence across the
system size at sufficiently low temperatures, whereas, at higher temperatures, coherence
is limited to smaller regimes. In this case, the system contains a quasi-condensate \cite{popov}.

The suppression of coherence with reduced dimensionality has been observed experimentally
in very elongated
3D Bose-Einstein Condensates (BECs) \cite{Arlt_3D,Klitzing,jan,Orsay_1D,Orsay_New}. In these systems,
the equilibrium coherence properties were found to be largely 
insensitive to density fluctuations, even 
at temperatures close to the critical temperature, where most of the atoms are
in the thermal cloud.
This can only be true, provided
phase fluctuations set in at temperatures much lower than density fluctuations,
which is typically considered as the only attainable experimental limit in such systems.

In this paper, we demonstrate 
that the
opposite regime, in which density fluctuations set in at a lower temperature than phase
fluctuations, is actually within current experimental reach. 
We determine optimum conditions
for reaching this novel regime for two atomic species, highlighting 
in particular the case of $^{23}$Na, which
appears to be a better candidate for such experiments.
The interplay between density and phase fluctuations is examined by comparing appropriately defined
characteristic temperatures for their respective onset.
Such an approach has already been used in other systems to examine the relative interplay
of phase and density fluctuations of the order parameter of the system.
A striking such example is the case of high-$T_{c}$ superconductors, for which
the interplay  between amplitude and phase fluctuations
is believed to be related to the `pseudogap' phase \cite{SC_1}. The present study 
probes the entire `crossover' between the previously observed 
regime of dominant phase fluctuations, and the currently unexplored regime
with dominant density fluctuations in weakly-interacting 1D ultracold atomic Bose gases.
This is performed by
varying the number of atoms in the system (Fig. 1), in close analogy
to the well-known phase-amplitude-fluctuations interplay studied in high-$T_{c}$ superconductors
 as a function of doping \cite{SC_2}.

\begin{figure}[b]
\includegraphics[width=7.cm]{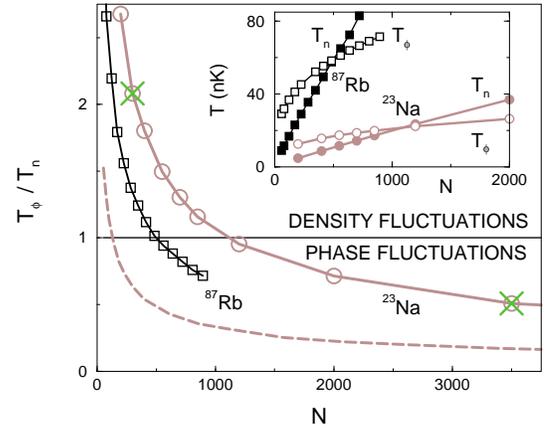}
\caption{
Ratio $T_{\phi}/T_{n}$ highlighting interplay between density and phase fluctuations,
as a function of atom number. Line $T_{\phi}=T_{n}$ indicates
crossover between
regimes of dominant phase ($T_{\phi}<T_n$)
and density fluctuations.
Circles: $^{23}$Na results for
$\omega_z = 2 \pi \times 10$ Hz, $\omega_\perp = 2 \pi \times 2500$ Hz. 
$^{87}$Rb  results also shown for same trap configuration (dashed line)
and for $\omega_z = 2 \pi \times 50$ Hz (squares)
Inset: Corresponding characteristic temperatures $T_{\phi}$ (open symbols) and $T_{n}$ (filled).
}
\end{figure}

In atomic gases, such characteristic temperatures can be determined as follows:
The temperature for phase fluctuations, $T_{\phi}$, is obtained by measuring the
normalized
off-diagonal one-body density matrix, also known in the
language of quantum optics
as the first order correlation function.
Roughly speaking, $T_{\phi}$ corresponds to the temperature at which the latter quantity decays
to zero at the edge of the quasi-condensate \cite{1d}, thus marking an approximate boundary 
between the regimes
of `true' condensation and quasi-condensation \cite{MFT}.
On the other hand, the temperature for density fluctuations can be determined
by measuring
the quasi-condensate fraction. 
In particular, we define $T_{n}$ as the temperature at which there is
a $10\%$ quasi-condensate depletion, since this can be readily measured experimentally.
For a given experimental configuration, both $T_{\phi}$ and $T_n$, and therefore their ratio, are fixed,
and experiments so far have only been performed in the limiting regime  $T_{\phi}/T_n \ll 1$ (see later).
However, the flexibility of atomic gas experiments enables complete tunability of the ratio  $T_{\phi}/T_n$, 
thus allowing one to probe the full
interplay between density and phase fluctuations.
This 
is explicitly shown in Fig. 1, for the particular example of variable 
atom number at fixed trap configuration.
As evident from this figure, the
novel regime of dominant density fluctuations, which can be approximately  
defined by $T_{\phi} > T_{n}$, is
attainable for a broad range of experimental parameters,
both for $^{23}$Na and $^{87}$Rb.
The corresponding characteristic temperatures
for both species are shown in the inset.

The equilibrium properties of the system are expected to depend on the interplay
between $T_{\phi}$ and $T_{n}$. 
In particular, the role of density fluctuations on the coherence of the gas can be determined by
looking at the difference in the measured coherence length of a depleted
quasi-condensate, from calculations based on
the corresponding undepleted system with the same {\em total} atom number.
Experiments to date have inferred a negligible effect of density fluctuations on the 
equilibrium coherence properties \cite{Arlt_3D,Klitzing,jan,Orsay_1D}, based on good agreement 
between measured data and a theory which
neglects density fluctuations \cite{1d,Gora_3d}. 
Surprisingly, this was argued to be the case even at temperatures
close to the effective `transition' temperature, where most of the quasi-condensate is 
known to be depleted by thermal excitations.
In this paper, we demonstrate that
such a conclusion is only correct in the regime
$T_{\phi} \ll T_{n}$, which is in fact the only regime explored experimentally to date.
Even within this regime, more recent work \cite{Orsay_New}
indicates a currently unaccounted for  reduction of 
$20\%$ in the coherence length, which we attribute to
density fluctuations.
We also demonstrate that this reduction can be largely enhanced by appropriate choice of parameters.

{\em Theory:}
Early theoretical work addressing trapped 1D systems
treated phase fluctuations exactly, but neglected density fluctuations \cite{1d}, thus
applicable only in the limit of an undepleted quasi-condensate.
Quasi-condensate depletion was subsequently included in a self-consistent
treatment of both phase and density fluctuations, leading us to the
construction of the 1D phase diagram in the weakly-interacting limit \cite{MFT}. 
The resulting theory was shown to be free of both infrared and
ultraviolet divergences and therefore valid in any dimension.
Importantly, and in stark contrast to other existing theories \cite{1d,Luxat,More}, 
such an equation of state enables
a direct {\em ab initio} determination of quasi-condensate density profiles.
In the harmonic trap, densities are obtained in the local density approximation,
the effective  temperature-dependent quasi-condensate size $R_{TF}(T)$ is obtained in the 
Thomas-Fermi limit, and the coherence of the system is investigated via 
the first-order normalized 
correlation function.

\begin{figure}[t]
\includegraphics[width=7.cm]{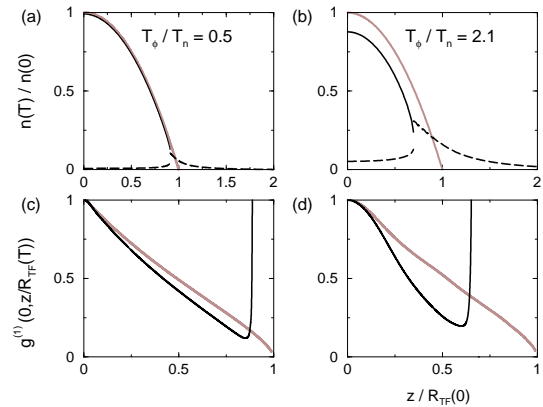}
\caption{
Top: 
Density profiles 
at $T=T_\phi$ for the opposite regimes of
$T_\phi/T_n=$ (a) $0.5$ and (b) $2.1$:
Quasi-condensate (solid black lines) and thermal (dashed black) density profiles
and corresponding zero-temperature profiles (grey) at fixed total atom
number ($N=$ (a) 3500, and (b) 300).
Densities scaled to the zero-temperature central density
and distance scaled to the zero-temperature
Thomas-Fermi radius $R_{TF}(0)$.
Bottom:
Corresponding normalized first-order spatial correlation functions 
$g^{(1)}(0,z/R_{TF}(0))$, evaluated at the trap centre, including
(black) and excluding (grey) quasi-condensate depletion.
Plotted data correspond to the highlighted $^{23}$Na points of Fig. 1, with
 $R_{TF}(T)/R_{TF}(0)=$ (a,c) 0.9 and (b,d) 0.7 and
quasi-condensate fractions $N_{0}/N=$ (a) 0.96, (b) 0.73.
}
\end{figure}

All quasi-condensate experiments performed to date \cite{Arlt_3D,Klitzing,jan,Orsay_1D,Orsay_New}
have been analyzed on the basis of the theory of
\cite{1d,Gora_3d} which entirely neglects both quantum and thermal fluctuations. Although the effect of
quantum fluctuations is small for experimentally relevant conditions, thermal fluctuations clearly
become important with increasing temperature, as evident from the experimentally observable quasi-condensate
depletion. It is therefore not a priori clear why such density fluctuations have not been observed
to couple to the phase fluctuations of the system. In order to ascertain the importance of density
fluctuations on the equilibrium properties of the system in an {\em ab initio} manner, we henceforth
compare the theory of \cite{MFT}, upon neglecting the quantum fluctuations, to the conventional
1D theory which additionally ignores thermal quasi-condensate depletion  \cite{1d,Luxat}.
The general expression for the first-order correlation function thus becomes
$ g^{(1)}(0,z) =
{\rm exp}\left( -\langle\left[\hat{\chi}(z)-\hat{\chi}(0)\right]^2 \rangle /2 \right),$
where  $\hat{\chi}(z)$ is the phase operator, satisfying \cite{MFT}
\begin{eqnarray}
\langle
\left[
\hat{\chi}(z)-\hat{\chi}(0)
\right]^2
\rangle & = & {4\pi \kappa l_z^4\over R_{\rm TF}^{3}(T)}
 \sum_{j=0}  2N(\hbar \omega_j) \nonumber \\ 
& & \hspace{-0.3cm}
\times \Bigg[
A_j^2\left(P_j(z/R_{\rm TF}(T))-P_j(0)\right)^2 \nonumber \\
& & \hspace{-0.3cm}
-B_j^2\left({P_j(z/R_{\rm TF}(T))\over 1-(z/R_{\rm TF}(T))^2}-P_j(0)\right)^2
\Bigg] \;. \nonumber
\end{eqnarray}  
Here $P_j(z)$ are Legendre polynomials of order $j$,
and $A_j=\sqrt{(j+1/2)\mu^\prime/\hbar\omega_j}$, 
$B_j=(\sqrt{(j+1/2)\hbar\omega_j/\mu^\prime})/2$. The
frequencies are given by $\omega_j=\sqrt{j(j+1)/2}\;\omega_z$,
$\mu'$ is the `renormalized' chemical potential including quasi-condensate depletion,
$\kappa$ the effective 1D scattering length and $l_{z}$ the harmonic oscillator length
corresponding to longitudinal confinement $\omega_{z}$.
$N(\hbar \omega_{j})$ is the usual Bose distribution function.
Density fluctuations are explicitely maintained in the above expression, via the temperature-dependent
quasi-condensate size $R_{TF}(T)$ appearing in the prefactor.
The general expression quoted above can be readily reduced to the
conventional theory \cite{1d,Luxat} which ignores quasi-condensate depletion
by replacing $R_{TF}(T)$ by the corresponding
zero temperature quasi-condensate size $R_{TF}(0)$ at the same {\em total} atom number,
and ignoring the $B_j$ contributions.
Since $R_{TF}(T) \propto \sqrt{\mu'}$, the former step is equivalent to replacing the renormalized
chemical potential at temperature $T$ by the corresponding zero-temperature one for the same total atom number.
Then, the `classical' approximation
$N(\hbar \omega_{j}) \approx k_B T / \hbar \omega_{j}$ leads to the definition of the characteristic temperature 
$T_{\phi}=(\hbar \omega_{z})^{2} N/k_{B} \mu$ \cite{1d}.
All presented results maintain the full
Bose distribution function $N(\hbar \omega_{j})=\left[{\rm exp}(\beta \hbar \omega_j) -1 \right]^{-1}$,
and are in the weakly-interacting regime, 
with $\gamma = 1/(n \xi)^{2} < 10^{-3}$ \cite{1d}.

{\em Results:}
The effect of density fluctuations on the equilibrium properties of the system 
is best assessed by investigating the latter
at the characteristic temperature, $T_{\phi}$, where phase fluctuations set in.
This study, shown in Fig. 2, reveals marked differences in both density profiles (top figures)
and correlation functions (bottom) between the regimes
of dominant phase and density fluctuations.
In particular, inclusion of density fluctuations leads to significant 
quasi-condensate depletion, reducing both the central
quasi-condensate density and the quasi-condensate size $R_{TF}(T) < R_{TF}(0)$. 
Moreover, within the region of validity of the finite temperature correlation functions, i.e.,
$z < R_{TF}(T)$,
density fluctuations are shown to lead to a potentially significant 
reduction in the coherence.

%
%

\begin{figure}[t]
\includegraphics[width=7.cm]{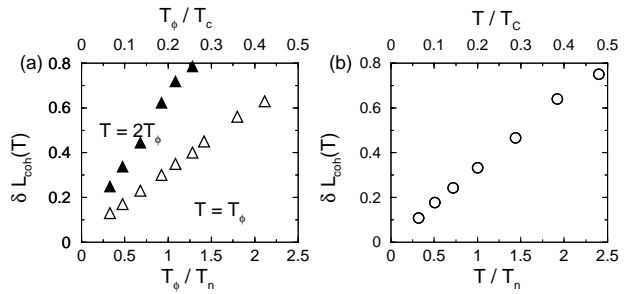}
\vspace{0.4cm}
\caption{
(a) Fractional deviation $\delta L_{coh}(T) = (L_{coh}^{0}(T) - L_{coh}^{T}(T))/L_{coh}^{T}(T)$
of the coherence length  $L_{coh}^{0}(T)$ ignoring density fluctuations,
from actual result $L_{coh}^{T}(T)$ including density fluctuations,
plotted as a function of the `crossover' parameter $T_\phi/T_n$ (bottom axis) and   $T_\phi/T_c$ (top).
Curves 
evaluated at $T=T_{\phi}$ (open triangles) and $T=2T_{\phi}$ (closed).
(b) Dependence of $\delta L_{coh}(T)$ on scaled temperature
$T/T_{n}$ (bottom), or equivalently $T/T_{c}$ (top).
Parameters as in Fig. 2.
}
\end{figure}

To quantify this reduction further, Fig. 3 shows the dependence of the
fractional deviation of the coherence length 
induced by the presence of density fluctuations
(a) on the `crossover' parameter
$T_{\phi}/T_n$, and (b) on appropriately scaled temperature.
The coherence length is
defined as the distance at which the correlation function decays to half its
value, i.e. $g^{(1)}(0,L_{coh}(T))=0.5$. 
The neglect of density fluctuations leads to a large overestimate of
the coherence length.
Importantly, this occurs
already at the crossover region $T_{\phi} \approx T_{n}$
even at temperatures as low as $T_{\phi}$
 at which density fluctuations are usually considered negligible \cite{1d}.
Using the approximate scaling
$T_n \approx 0.2 T_c$ valid for the examined parameters, we conclude from Fig. 3(a) that
density fluctuations should be
observable at very low temperatures, even within experimental uncertainties, 
provided  $T_{\phi}/T_c>0.1$.
The approximately linear increase in temperature of $\delta L_{coh}(T)$ plotted in
Fig. 3(b) is universal when temperatures are scaled in terms of
$T_n$, or equivalently, $T_{c}$. 
However, corresponding features 
differ {\em noticeably} in terms of $T/T_{\phi}$,
suggesting that $T_{\phi}$ is not the most appropriate scaling parameter
for such studies.
%

The dependence of the coherence length, scaled to the zero-temperature system size, 
on appropriately scaled temperature is shown in
Fig. 4.
Filled (open) 
symbols correspond to the cases  $T_{\phi} > T_{n}$ ($T_{\phi} < T_{n}$), with (black)
or without (grey) density fluctuations included. 
The coherence length decreases approximately exponentially with temperature, 
consistent with experimental findings \cite{Orsay_1D}.
Importantly, however, {\em in both cases},
 density fluctuations lead to a considerable shift of the curves to lower values.
This is also true when temperatures are scaled in terms of $T_{\phi}$ (Fig. 4(b)),
as usually done experimentally 
\cite{Arlt_3D,Klitzing,jan,Orsay_1D,Orsay_New},
in which case
results {\em without} density fluctuations fall roughly on a universal curve
(inset). 
However, inclusion of density fluctuations is found to lead to
a noticeable downward shift of the correlation function,
already in the regime $T_{\phi}<T_{n}/2$ (black circles).
The magnitude of this shift clearly increases with increasing $T_{\phi}/T_n$, as demonstrated by the
black squares. 
Note that results with density fluctuations are fully consistent with
corresponding results based on a stochastic Langevin approach in the classical field approximation \cite{Stoof_Noisy}, 
to be presented elsewhere.

{\em Current Experiments:}
Measurements of phase fluctuations to date have only been performed close to the 1D regime. 
In these experiments, the departure from pure 1D excitations arising from $\mu \sim$ few $\hbar \omega_{\perp}$
and the small values of the phase degeneracy temperature ($T_{\phi}/T_c \ll 0.1$) tend to suppress
density fluctuations.
Early experiments based on  interferometric techniques \cite{jan}, or Bragg spectroscopy \cite{Orsay_1D}
found no evidence of density fluctuations on the equilibrium 
coherence properties, within their observational uncertainties.
Nonetheless,
more recent measurements of the temperature dependence of the scaled coherence length
performed 
at temperatures $T/T_{\phi}<6$ \cite{Orsay_New}
cannot be interpreted by a 3D theory which ignores density fluctuations
\cite{Gora_3d} unless the corresponding predictions are
shifted to lower values by about $20\%$. 
We believe that this shift arises, at least partly, from the effect of density fluctuations, as supported
qualitatively by our present 1D analysis.


\begin{figure}[t]
\includegraphics[width=9.cm]{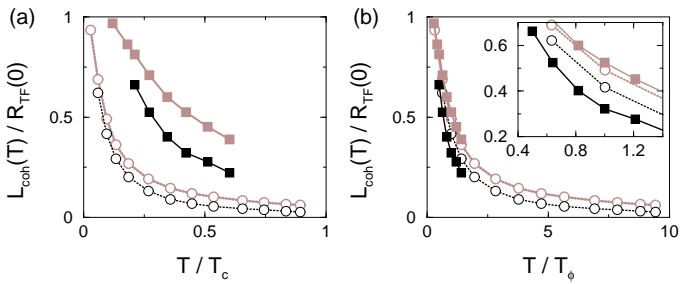}
\caption{
Temperature dependence of the scaled coherence length 
$L_{coh}(T)/R_{TF}(0)$ on (a) $T/T_c$ and (b) $T/T_\phi$
for parameters of Fig. 2:
 $T_\phi/T_c \approx$ 0.1 (open symbols) and 0.4 (filled),
without (grey) or with (black) density fluctuations.
Inset: Deviation in regime
 $0.4<T/T_{\phi}<1.4$. 
}
\end{figure}

{\em Proposed Experiment:}
The crossover between the regimes of dominant phase and density fluctuations
can be probed in a controlled manner
by repeated measurements of the equilibrium coherence properties
of the system at fixed temperature
and variable atom number.
Instead of decreasing the atom number, one could achieve the same effect by
reducing
the scattering length via a Feshbach resonance \cite{Feshbach},
since both these effects decrease the system nonlinearity.
The `crossover' parameter determining the relative interplay between phase and density 
fluctuations was numerically found to obey 
$T_{\phi}/T_n \approx \eta (\hbar \omega_{z} / \mu)$.
We first focus our discussion on $^{23}$Na, which appears to be a good candidate for such experiments,
and for which $\eta \approx 25$. 
Based on the usual 3D scattering length of 2.75nm, this
gives
$T_{\phi}/T_n \approx 10^{4} {\rm Hz}^{2/3} \times (\omega_{z}/\omega_{\perp}^{2})^{1/3} / N^{2/3}$.

The proposed regime is achieved when a number of competing conditions are satisfied:
Firstly, 1D thermal excitations require $kT < \hbar \omega_{\perp}$,
implying a relatively large transverse confinement of the order
of a few kHz, which is easily achievable in experiments with atom chips \cite{AtomChip}. 
The condition for 1D quasi-condensation, $\mu < \hbar \omega_{\perp}$,
translates into a condition for the maximum number of atoms at given trap confinement,
namely $N^{2} (\omega_{z}^{2}/\omega_{\perp}) < C_{1}$.
Finally, the requirement for dominant density fluctuations, $T_{\phi} > T_{n}$, places a
competing condition $N^{2} (\omega_{\perp}^{2}/\omega_{z}) < C_{2}$.
For $^{23}$Na,  we find $C_{1} \approx {\rm few} \times 10^{8}$Hz 
and $C_{2} \approx {\rm few} \times 10^{12}$Hz.
At the chosen confinement  $\omega_{z} =2 \pi \times 10$Hz 
and $\omega_{\perp}=2 \pi \times 2.5$kHz, the 1D quasi-condensate condition is the most restrictive
for the maximum number of atoms required, whereas
the opposite
is true for the same aspect ratio $\omega_{\perp}/\omega_{z}$ but larger 
$\omega_{\perp}=2 \pi \times 10$kHz.

Throughout this work,
we have focused on $^{23}$Na,
on the grounds that such experiments may be easier
because $^{23}$Na enters the regime of dominant density fluctuations
with a relatively large atom number
$N >1000$ for a broad parameter range.
However, this comes at the
expense of a stringent requirement on low temperatures, typically around $20$nK.
The same regime is also attainable with $^{87}$Rb, for which $\eta \approx 10$.
Although $^{87}$Rb requires
lower aspect ratios and higher temperatures than $^{23}$Na, 
it nonetheless requires a reduced number of atoms, with $N<700$ even for optimized aspect ratios.
The optimum candidate for such an experiment 
thus depends on the flexibility of the experimental set-up,
with respect to cooling efficiency and atom detection techniques.

The regime of dominant density fluctuations typically requires the phase fluctuation temperature,
$T_{\phi}$,
to be comparable to the effective `transition' temperature $T_c$.
In this regime, the suppressed phase fluctuations are dominated
by their coupling to density fluctuations.
Such a regime  may
prove useful for potential applications, due to
the presence of large coherence even at relatively high temperatures $T \sim 0.5 T_{c}$.
However, the effect of density fluctuations on the coherence length
can already be observed at
the much weaker approximate condition $T_{\phi} > 0.1 T_{n}$,
which increases the value
$C_{2}$ by a factor of 10.

In conclusion, we proposed a new regime of quasi-one-dimensional weakly-interacting atomic gases,
in which density fluctuations set in at a lower temperature than phase fluctuations.
This regime was shown to be experimentally accessible for small atom numbers, low temperatures
and moderate aspect ratios.
It
can be experimentally identified by the noticeable
reduction in the equilibrium coherence length of the quasi-condensate,
whose extent
depends critically on the ratio of the phase degeneracy
temperature $T_{\phi}$ to the effective transition temperature $T_{c}$.
%
Given the pronounced nature of density fluctuations in homogeneous systems, an alternative observation
of dominant density fluctuations can be performed in the new box-like trap \cite{Box}.
The unprecedented experimental control offered by ultracold atomic gases may offer
further insight into the 
interplay between phase and density fluctuations in
analogous condensed-matter systems.


I am indebted to Henk Stoof for numerous insightful discussions and to David Luxat for early collaboration
on this project.
Discussions with C. de Morais Smith and U. Al Khawaja are also acknowledged.
This work was supported by the Nederlandse Organisatie voor Wetenschaplijk Onderzoek (NWO).

\end{document}